\documentclass[print]{article}
\usepackage{amsmath,amssymb}
\usepackage{graphicx}

\begin{document}

\begin{center}\large \bf  Dynamical Symmetry and Quantum
Information Processing with Electromagnetically Induced
Transparency\end{center}

\begin{center}Xiong-Jun Liu$^{a,b}$\footnote{Electronic address:xiongjunliu@yahoo.com.cn},
Hui Jing$^c$, Xin Liu$^{a,b}$ and Mo-Lin Ge$^{a,b}$\end{center}

\begin{center}a. Theoretical Physics Division, Nankai Institute of
Mathematics,Nankai University, Tianjin 300071, P.R.China\\
 b. Liuhui Center for Applied Mathematics, Nankai
University and Tianjin University, Tianjin 300071, P.R.China\\
 c. State Key Laboratory of Magnetic Resonance and Atomic and Molecular
 Physics,\\
Wuhan Institute of Physics and Mathematics, CAS, Wuhan 430071, P.
R. China\end{center}

\begin{abstract}
We study in detail the interesting dynamical symmetry and its
applications in general many-level and many-ensemble atomic
systems with electromagnetically induced transparency (EIT). By
discovering the symmetrical Lie group of various atomic systems,
the novel applications to quantum memory and quantum entanglement
between photons or atomic ensembles are investigated. \\

PACS numbers: 03.67.-a, 03.65.Fd, 03.67.Mn, 42.50.Gy

\end{abstract}

\baselineskip=16pt


\section{Introduction}
\indent During the last decade or so, rapid advances have been
witnessed in both experimental and theoretical aspects towards
probing the novel mechanism of Electromagnetically Induced
Transparency (EIT) \cite{2} and its many potential applications
\cite{3,4,5}. In particular, based on the elegant "dark-state
polaritons" (DSPs) theory proposed by Fleischhauer and Lukin
\cite{6},  the quantum memory techniques are now actively explored
by exchanging the quantum state information between the quantized
light field and the metastable collective atomic field \cite{7}.
DSP is a new quantum field which is the superposition of the light
field amplitude and the atom coherence between two lower levels of
the $\Lambda$ type $three$-level atoms, and it describes the total
system of the optical and collective atomic fields. In linear
theory where the two-photon detuning of the light pulses is zero,
the dynamical evolution of DSPs can lead to a perfect state
mapping from the photonic branch into the atomic excitation one
and vice versa by adiabatically adjusting the coupling laser
\cite{6,7}.

The dynamical symmetry of multi-level atomic system interacting
with light fields was studied by D. A. Lidar et al. \cite{Lidar}.
On other hand, a semidirect product group in $three$-level atomic
system under the condition of larger atom number and low
collective excitation limit \cite{6} with EIT was discovered by
Sun et al \cite{8}, and the the validity of adiabatic passage
condition for the dark states is also investigated in this
technique. After that, a series research on the study of hidden
symmetry as well as its application to quantum information with
$four$-level atomic system and many atomic ensembles were done
recently \cite{liu,atom}. All these works indicate many
interesting hidden symmetrical properties in various atomic
systems with EIT. In this paper, by discovering the symmetrical
Lie group, we examine in detail the general definition of
dark-state polariton (DSP) operators, and then the dark-states in
different atomic systems. Also, it is interesting to find that the
symmetrical properties of the multi-level system and
multi-atomic-ensemble system are dependent on some characteristic
parameters such as the coupling constant $g_i$ and Rabi frequency
$\Omega_i$ etc.. Furthermore, a controllable scheme to generate
quantum entanglement between atoms or lights via quantized DSPs
theory is discussed, which might be experimentally implemented in
the near future..

The development herein is outlined as follows. In section II, we
discuss the dynamical symmetry by discovering the Lie algebra
structure of various atomic systems including multi-level and
multi-atomic ensembles cases etc.; In section III, we respectively
examine the general definition of DSP operators and then quantum
memory for photons via DSP theory of these systems; Generation of
different formalisms of entanglement between atoms or lights via
quantized DSPs theory are discussed in section IV; In the last
section, we conclude and further discuss the dynamical symmetry
and the applications in these EIT-systems.

\section{Hidden symmetrical group in electromagnetically induced transparency}
\subsection{Complex $m$-level ($m>3$, multi-level) atomic system}

The system we consider is shown in Fig. 1 (a), a collection of $N$
double $\Lambda$ type $m$-level ($m\geq3$, multi-level) atoms
interact with $m-2$ single-mode quantized fields which couple the
transitions from the ground state $|b\rangle$ to excited state
$|e_{\sigma}\rangle$ $(1\leq \sigma\leq m-2)$ with coupling
constants $g_{\sigma}$, and $m-2$ classical control ones, which
couple the transitions from the metastable state $|c\rangle$ to
excited one $|e_{\sigma}\rangle$ with time-dependent
Rabi-frequencies $\Omega_{\sigma}(t)$. Generalization to
multi-mode probe pulse case is straightforward. Considering all
transitions at resonance, the interaction Hamiltonian of the total
system can be written as:

\begin{figure}[ht]
\center{\includegraphics[width=0.8\columnwidth]{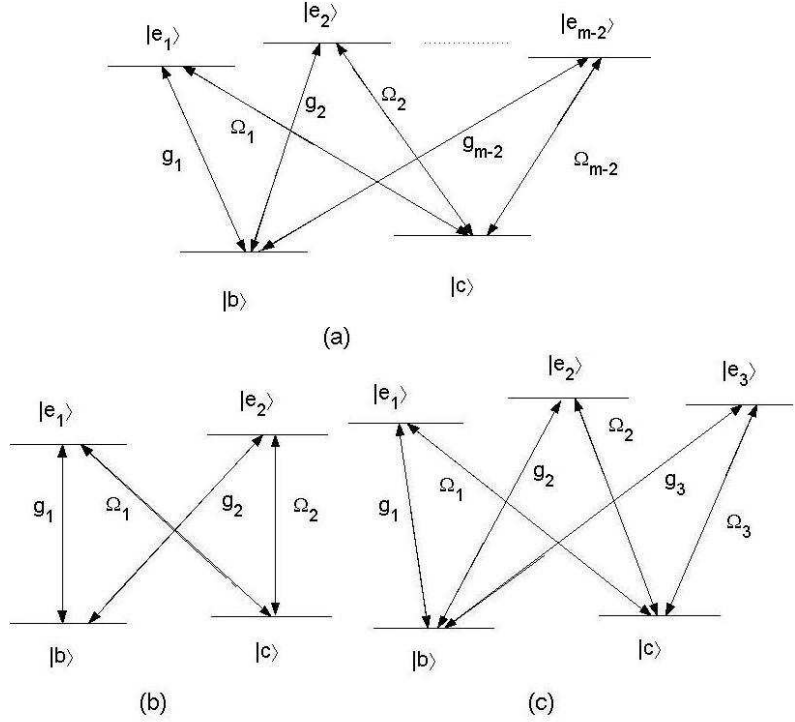}
\caption{Single ensemble composed of multi-level atoms interacts
with with many single-mode quantized and classical control fields.
In particular, (a) General $m$-level atomic-ensemble; (b)
Four-level case; (c) Five-level $W$ type atom-ensemble
case.}}\label{}
\end{figure}

\begin{eqnarray}\label{eqn:1}
\hat H=\sum_{\sigma=1}^{m-2}g_{\sigma}\sqrt{N}\hat a_{\sigma}\hat
E_{\sigma}^{\dag}+\sum_{\sigma=1}^{m-2}\Omega_{\sigma}\hat
T_{e_{\sigma}c}+h.c.,
\end{eqnarray}
where subscription $\sigma$ denotes the corresponding excited
state and the collective atomic excitation operators:
\begin{equation}\label{eqn:2}
\hat
E_{\sigma}=\frac{1}{\sqrt{N}}\sum_{j=1}^{N}\hat\sigma_{be_{\sigma}}^{j},
\ \hat C=\frac{1}{\sqrt{N}}\sum_{j=1}^N\hat\sigma_{bc}^{j},
\end{equation}
with $\hat\sigma^j_{\mu\nu}=|\mu\rangle_{jj}\langle\nu|
(\mu,\nu=b,c,e_1,e_2,...,e_{m-2})$ being the flip operators of the
$j$-th atom between states $|\mu\rangle$ and $|\nu\rangle$, and
\begin{equation}\label{eqn:3}
\hat T^{-}_{\mu\nu}=\hat T_{\mu\nu}=\sum_{j=1}^{N}\hat\sigma
_{\mu\nu}^{j}, \ \ \hat T_{\mu\nu}^{+}=(\hat
T^{-}_{\mu\nu})^{\dagger },
\end{equation}
where $\mu\neq\nu=c,e_1,e_2,...,e_{m-2}$. Denoting by \cite{dick}
$|b\rangle=|b_1,b_2,...,b_N\rangle$ the collective ground state
with all $N$ atoms staying in the same single particle ground
state $|b\rangle$, we can easily give other quasi-spin wave states
by the operators defined in formula (\ref{eqn:2}):
$|e_{\sigma}^n\rangle=[n!]^{-1/2}(\hat
E_{\sigma}^{\dag})^n|b\rangle$ and $|c^n\rangle=[n!]^{-1/2}(\hat
C^{\dag})^n|b\rangle$. For the EIT case, we consider two
approximation conditions \cite{6,7}: i) The system include a very
large number of atoms, i.e. $N\gg1$; ii) The low atomic excitation
condition, i.e. the control fields are much stronger than the
quantized probe fields and only a few atoms occupy the metastable
state $|c\rangle$ and excited states $|e_j\rangle$. It then
follows that $[\hat E_i,\hat E_j^{\dag}]=\delta_{ij}$ and $[\hat
C,\hat C^{\dag}]=1$ and all the other commutators are zero, which
shows the mutual independence between these bosonic operators
$\hat E_{\sigma}$ and $\hat C$. On the other hand, the $m^2-3m+2$
collective operators $\hat T_{\mu\nu}$ satisfy the $u(m-1)$
commutation relation: $ [\hat T_{\alpha\beta},\hat
T_{\mu\nu}]=\delta^{\beta\mu}\hat
T_{\alpha\nu}-\delta^{\alpha\nu}\hat T_{\mu\beta} $. Thus the
operators ($\hat T^{\pm}_{\mu\nu}, \hat
T_{\mu\nu}^z$)($\mu,\nu=c,e_1,e_2,...,e_{m-2}$) compose the
$m^2-2m+1$ generators of the algebra $su(m-1)$, here
\begin{equation}\label{eqn:4}
\hat T_{\mu\nu}^{z}=\sum_{j=1}^{N}(\hat\sigma
_{\mu\mu}^{j}-\hat\sigma _{\nu\nu}^{j})/2, \
(\mu\neq\nu=c,e_1,e_2,...,e_{m-2}),
\end{equation}
with the relation $\hat T_{\mu\nu}^{z}=\hat T_{\mu\rho}^{z}-\hat
T_{\rho\nu}^{z}$. Considering $[\hat T_{ce_{\sigma}}^{+},\hat
E_{\sigma}]=-\hat C$, $[\hat T_{ce_{\sigma}}^{-},\hat C]=-\hat
E_{\sigma}$, $[\hat T_{e_ie_j}^{+},\hat E_k]=\delta_{jk}\hat
E_i-\delta_{ik}\hat E_j$, $[\hat T_{e_ie_j}^{-},\hat
E_k]=\delta_{ik}\hat E_j-\delta_{jk}\hat E_i$ and denoting by
$h_{m-1}$ the algebra generated by $(\hat E_{\sigma},\hat
E^{\dag}_{\sigma},\hat C,\hat C^{\dag})$, we then obtain
$[su(m-1),h_{m-1}]\subset h_{m-1}$ which means that the dynamical
symmetry of the $m$-level atomic system is governed by a
semidirect product Lie group \cite{group}
$SU(m-1)\overline{\otimes}H_{m-1}$ in large $N$ limit and low
excitation condition. In general, the dynamical symmetry of a
$m$-level atomic system is governed by $SU(m)$ \cite{Lidar}, e.g.
the Gell-Mann dynamical symmetry $SU(3)$ for three-level quantum
system \cite{gell}. However, here in the large atom number limit
and low excitation condition, the dynamical symmetry of the
multi-level EIT system is governed by a semi-direct Lie group.
Particularly, when $m=3$, i.e., for the usual $three$-level
system, the dynamical symmetry is governed by the simplest
$SU(2)\overline{\otimes}H_2$ group \cite{8}, while the
$four$-level double $\Lambda$ ($m=4$) system \cite{four1,four2} is
governed by $SU(3)\overline{\otimes}H_{3}$ \cite{liu} (Fig. 1(b))
and the $five$-level $W$-type system governed by
$SU(4)\overline{\otimes}H_4$ (Fig. 1(c)), etc..

\subsection{Multi-atomic-ensemble system of $three$-level atoms}

In this subsection we consider a cloud of identical atoms with the
$three$-level $\Lambda$ type structure which is shown in Fig. 2.
Atoms of the $l$-th $(l=1,2,...k)$ atomic ensemble interact with
the input single-mode quantized field with coupling constants
$g_l$, and one classical control filed with time-dependent
Rabi-frequencies $\Omega_l(t)$. Considering all transitions at
resonance, the interaction Hamiltonian of the total system can be
written as:

\begin{center}\begin{figure}[ht]
\center{\includegraphics[width=0.7\columnwidth]{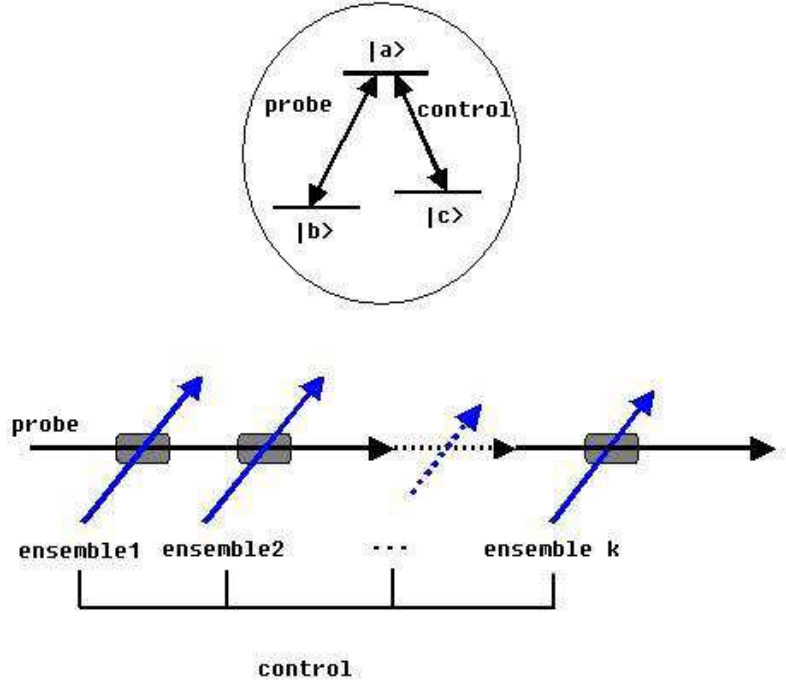}
\caption{(color online) EIT process for many ensembles composed of
$\Lambda$ type atoms located in the straight-line
configuration.}}\label{}
\end{figure}\end{center}

\begin{eqnarray}\label{eqn:6}
\hat H=\sum_{\sigma=1}^kg_{\sigma}\sqrt{N_{\sigma}}\hat a\hat
A_{\sigma}^{\dag}+\sum_{\sigma=1}^k\Omega_{\sigma}(t)\hat
T^+_{\sigma}+h.c.,
\end{eqnarray}
where the subscript $\sigma$ denotes the corresponding atomic
ensemble and the collective atomic excitation operators:
\begin{equation}\label{eqn:7}
\hat
A_{\sigma}=\frac{1}{\sqrt{N_{\sigma}}}\sum_{j=1}^{N_{\sigma}}e^{-i\bf
k_{ba}\cdot\bf r^{(\sigma)}_j}\hat\sigma_{ba}^{j(\sigma)}, \ \
\hat
C_{\sigma}=\frac{1}{\sqrt{N_{\sigma}}}\sum_{j=1}^{N_{\sigma}}e^{-i\bf
k_{bc}\cdot\bf r^{(\sigma)}_j}\hat\sigma_{bc}^{j(\sigma)}, \ \
\sigma=1,2,...,k
\end{equation}
with $\hat\sigma^i_{\mu\nu}=|\mu\rangle_{ii}\langle\nu|
(\mu,\nu=a,b,c)$ being the flip operators of the $i$-th atom
between states $|\mu\rangle$ and $|\nu\rangle$, $\bf k_{ba}$ and
$\bf k_{ca}$ are, respectively, the wave vectors of the quantum
and classical light fields, $\bf k_{bc}=\bf k_{ba}-\bf k_{ca}$ and
\begin{equation}\label{eqn:8}
\hat T^{-}_{\sigma}=(\hat
T_{\sigma}^{+})^{\dagger}=\sum_{j=1}^{N_{\sigma}}e^{-i\bf
k_{ca}\cdot\bf r^{(\sigma)}_j}\hat\sigma _{ca}^{j(1)}.
\end{equation}
Denoting by
$|b^{({\sigma})}\rangle=|b^{({\sigma})}_1,b^{({\sigma})}_2,...,b^{({\sigma})}_{N_{\sigma}}\rangle
({\sigma}=1,2,...,k)$ the collective ground state of the
${\sigma}$-th atomic ensemble with all atoms staying in the same
single particle ground state $|b\rangle$, we can easily give other
quasi-spin wave states by the operators defined in formula
(\ref{eqn:11}): $|a^n_{(\sigma)}\rangle=[n!]^{-1/2}(\hat
A_{\sigma}^{\dag})^n|b^{(\sigma)}\rangle$ and
$|c^n_{(\sigma)}\rangle=[n!]^{-1/2}(\hat
C_{\sigma}^{\dag})^n|b^{(\sigma)}\rangle$. Similarly, in large
$N_{\sigma}$ limit and low excitation condition, it follows that
$[\hat A_{(i)},\hat A^{\dag}_{(j)}]=\delta_{ij}, [\hat
C_{(i)},\hat C^{\dag}_{(j)}]=\delta_{ij}$ and all the other
commutators are zero, which shows the mutual independence between
these bosonic operators $\hat A_{i}$ and $\hat C_{i}$. On the
other hand, one can easily find the commutation relations: $[\hat
T^+_{i},\hat T^-_{j}]=\delta_{ij}\hat T^z_{j}$ and $[\hat
T^z_{i},\hat T^{\pm}_{j}]=\pm\delta_{ij}\hat T^{\pm}_{j}$, where
\begin{equation}\label{eqn:9}
\hat T_{\sigma}^{z}=\sum_{j=1}^{N_{\sigma}}(e^{-i\bf
k_{aa}\cdot\bf r^{(\sigma)}_j}\hat\sigma
_{aa}^{j(\sigma)}-e^{-i\bf k_{cc}\cdot\bf
r^{(\sigma)}_j}\hat\sigma _{cc}^{j(\sigma)})/2 \
(\sigma=1,2,...,m)
\end{equation}
are two traceless operators. Thus the operators $(\hat
T^{\pm}_{\sigma}, \hat T^z_{\sigma})$ generate the
$\oplus_{\sigma}su(2)$ algebra. Considering $[\hat T_{i}^{+},\hat
A_{j}]=-\delta_{ij}\hat C_{j}$, $[\hat T_{i}^{-},\hat
C_{j}]=-\delta_{ij}\hat A_{j}$ and denoting by $h_{2m}$ the
Heisenberg algebra generated by $(\hat A_i,\hat A_i^{\dag},\hat
C_i,\hat C_i^{\dag}; i=1,2,...,k)$, we then obtain
$[\oplus_{\sigma}su(2),h_{2k}]\subset h_{2k}$ which means that the
dynamical symmetry of the double $\Lambda$ system is governed by a
semidirect product Lie group \cite{group}
$(\otimes_{\sigma}SU(2))\overline{\otimes}H_{2k}$ in large
$N_{\sigma}$ limit and low excitation condition. In particular,
for $k=2$, the symmetrical group reads
$SO(4))\overline{\otimes}H_{4}$ \cite{atom}.

\section{Quantum memory process in multi-level and multi-ensemble atomic system}
The discovery of dynamical symmetry in above section leads us, by
the spectrum generating algebra method \cite{group}, to find
$H-$invariant subspaces, in which one can diagonalize the
Hamiltonian easily. As is known, the zero-eigenvalue subspace
composed of dark states is the key definition in quantum memory
with EIT technique \cite{6,7,Lidar,8}. During the quantum memory
process when the quantum states are adiabatically transferred from
lights to collective atom coherence, the total system of the atoms
and quantized probe light should be restricted in the dark-state
subspace \cite{6,7,Lidar,8}, therefore the key point of studying
this process is to obtain the dark states of the total system. On
the other hand, the dark states can be generated by the dark-state
polaritons (DSPs) operator which commutes with the Hamiltonian
operator \cite{6,7,8}, so we firstly study the general definition
of the DSPs operator in the general multi-level atomic and
multi-ensemble atomic systems, and can then easily study the
quantum memory process by generating the dark states of the
system.

DSPs operator can be constructed basing on two properties: 1) It
commutes with the Hamiltonian and satisfies the bosonic
commutation relation; 2) It is the superposition of the collective
atomic excitation operator $\hat C$ and the annihilation operators
of the quantized probe lights. For this the dark-state subspace is
a collection composed of zero-eigenstates excluding any excited
state $|e_j\rangle$ (in multi-level system) or $|a\rangle$ (in
multi-atomic-ensemble system). For the derivation of the DSPs
operator, firstly we can obtain its form of the three-level Lambda
system \cite{8}, and four-level double Lambda system \cite{liu},
and also the five-level case. Then, by induction we can obtain the
general definition of the DSPs operator in m-level case, which is
similar to that in multi-atomic-ensemble system.

\subsection{Quantum memory with a $m$-level atomic system}
Firstly, we study the general definition of DSPs of the
single-atomic-ensemble system with many $m$-level atoms. Based on
the above analysis of the properties of DSP operator, the new type
of dark-state-polaritons operator of the $m$-level system can be
defined as
\begin{equation}\label{eqn:d1}
\hat d=\cos\theta\prod_{j=1}^{m-3}\cos\phi_j\hat
a_1+\cos\theta\sum_{l=2}^{m-2}\sin\phi_{l-1}\prod_{j=l}^{m-3}\cos\phi_j\hat
a_l-\sin\theta\hat C,
\end{equation}
where the mixing angles $\theta$ and $\phi_j$ are defined through
\begin{equation}\label{eqn:10}
\tan\theta=\frac{g_1g_2...g_{m-2}\sqrt{N}}{\bigr[\sum_{j=1}^{m-2}\bigr(\Omega_j^2\prod_{l=1,l\neq
j}^{m-2}g_l^2\bigr)\bigr]^{1/2}}
\end{equation}
and
\begin{equation}\label{eqn:11}
\tan\phi_j=\frac{\prod_{l=1}^{j}g_l\Omega_{j+1}}{\bigr[\sum_{l=1}^{j}\bigr(\Omega^2_l\prod_{s=1,s\neq
l}^{j+1}g^2_s\bigr)\bigr]^{1/2}}.
\end{equation}

The eq.(\ref{eqn:11}) provides us
$\tan\phi_1=g_1\Omega_2/g_2\Omega_1,
\tan\phi_2=g_1g_2\Omega_3/\sqrt{\Omega_1^2g_2^2g_3^2+\Omega_2^2g_1^2g_3^2}
...$, etc. By a straightforward calculation one can verify that
\begin{equation}\label{eqn:com1}
[\hat d,\hat d^{\dag}]=1, \ \ \ \  [\hat H,\hat d \ ]=0,
\end{equation}
hence the general atomic dark states of $m$-level system can be
obtained through $|D_n\rangle=[n!]^{-1/2}(\hat
d^{\dag})^n|0\rangle$, where
\begin{equation}\label{eqn:20}
|0\rangle=\underbrace{|0, 0,...,
0}_{m-2}\rangle_{photon}\otimes|b\rangle_{atom}
\end{equation}
are the collective ground states \cite{ground} with
$|0,0,...,0\rangle_{photon}$ denoting the electromagnetic vacuum
of $m-2$ quantized probe fields.

Based on the above result we here discuss a novel phenomenon.
Initially, only one weak probe light (described by the coherent
state $|\alpha_1\rangle$ with $\alpha_1=\alpha_0$) is injected
into the atomic ensemble to couple the transition from $|b\rangle$
to $|e_1\rangle$, one strong control field is used to couple the
transition from $|c\rangle$ to $|e_1\rangle$ and all other light
fields ($m-3$ probe fields and $m-3$ control fields) are off. For
this the mixing angles $\theta=0$, $\phi_j=0$ and the initial
total state of the quantized field and atomic ensemble reads $
|\Psi_0\rangle=\sum_{n}P_n(\alpha_0)|n,\underbrace{0,0,...,0}_{m-3}
\rangle_{photon}\otimes|b\rangle_{atom} $, where
$P_n(\alpha_0)=\frac{\alpha_0^n}{\sqrt{n!}}e^{-|\alpha_0|^2/2}$ is
the probability of distribution function. Subsequently, the mixing
angle $\theta$ is adiabatically rotated to $\pi/2$ by turning the
control field off, and the quantum states of the probe light
$|\alpha_1\rangle$ is fully mapped into the collective atomic
excitations, i.e.
$|\Psi_t\rangle=\sum_{n}P_n(\alpha_0)|\underbrace{0,0,...,0}_{m-2}
\rangle_{photon}\otimes|c^n\rangle_{atom}$. Finally, when all
$m-2$ control fields are all turned back on and the mixing angle
$\theta$ is rotated back to $\theta=0$ again with $\phi_j$ to some
value $\phi_{ej}$ which are only determined by the
Rabi-frequencies of the re-applied control fields, we finally
obtain
\begin{eqnarray}\label{eqn:split1}
|\Psi_e\rangle&=&\sum_{n}P_n(\alpha_0)|D_n(\theta=0)\rangle\nonumber\\
&=&\sum_{j}\sum_{l}...\sum_{f}P_j(\alpha_{e1})P_l(\alpha_{e2})...P_f(\alpha_{e(m-2)})
|b\rangle\otimes|j,l,...,f\rangle\nonumber\\
&=&|b\rangle_{atom}\otimes|\alpha_{e1},\alpha_{e2},...,\alpha_{e(m-2)}\rangle_{photon},
\end{eqnarray}
where $\alpha_{e1}=\alpha_0\prod_{j=1}^{m-3}\cos\phi_{ej}$ and
$\alpha_{el}=\alpha_0\sin\phi_{e(l-1)}\prod_{j=l}^{m-3}\cos\phi_{ej},
(l=2,3,...,m-2)$ are the parameters of the released coherent
lights. The above expression clearly shows that the injected
quantized field can convert into $m-2$ different coherent pulses
$|\alpha_{ej}\rangle (j=1,2,...,m-2)$ after a proper evolution
manipulated by the control fields. Particularly, if the strengths
of all re-applied control fields equal each other, the output
probe lights read
$\alpha_{e1}=\alpha_{e2}=...=\alpha_{e(m-2)}=\alpha_0/\sqrt{m-2}$.
Obviously, this novel mechanism can be extended to other cases of
the injected field, say, in presence of a non-classical or
squeezed light beam.

\subsection{Quantum memory with k-atomic-ensemble system}
Now, to give a clear description of the interesting quantum memory
process in this $k$-atomic-ensemble system composed of $\Lambda$
type $three$-level-atoms, we define the new type of
dark-state-polaritons operator as
\begin{equation}\label{eqn:d2}
\hat d=\cos\theta\hat a-\sin\theta\prod_{j=1}^{k-1}\cos\phi_j\hat
C_1-\sin\theta\sum_{l=2}^k\sin\phi_{l-1}\prod_{j=l}^{k-1}\cos\phi_j\hat
C_l,
\end{equation}
where the mixing angles $\theta$ and $\phi_j$ are defined through
\begin{equation}\label{eqn:12}
\tan\theta=\frac{\bigr[\sum_{j=1}^{k}\bigr(g_j^2N_j\prod_{l=1,l\neq
j}^{k}\Omega^2_l\bigr)\bigr]^{1/2}}{\Omega_1\Omega_2...\Omega_k}
\end{equation}
and
\begin{equation}\label{eqn:13}
\tan\phi_j=\frac{g_{j+1}\sqrt{N_{j+1}}\prod_{l=1}^{j}\Omega_l}{\bigr[\sum_{l=1}^{j}\bigr(g^2_lN_l\prod_{s=1,s\neq
l}^{j+1}\Omega^2_s\bigr]^{1/2}},
\end{equation}
where one finds
$\tan\phi_1=g_2\sqrt{N_2}\Omega_1/g_1\sqrt{N_1}\Omega_2,
\tan\phi_2=g_3\sqrt{N_3}\Omega_1\Omega_2/\sqrt{g_1^2N_1\Omega_2^2\Omega_3^2+g_2^2N_2\Omega_1^2\Omega_3^2}$,
etc. Also, by a straightforward calculation one can verify that
$[\hat d,\hat d^{\dag}]=1$ and $ [\hat H,\hat d \ ]=0 $, hence the
general atomic dark states can be obtained through
$|D_n\rangle=[n!]^{-1/2}(\hat d^{\dag})^n|0\rangle$, where
$|0\rangle=|b^{(1)}, b^{(2)},...,
b^{(k)}\rangle_{atom}\otimes|0\rangle_{photon}$ and
$|0\rangle_{photon}$ denotes the electromagnetic vacuum of the
quantized probe field.

Similar to the discussion in above subsection, we can investigate
the quantum memory process in the multi-ensemble atomic system.
Initially the total state reads (meanwhile $\theta=0$ or the
external control fields are very strong):
$|\Psi_0\rangle=\sum_{n}P_n(\alpha_0)|b^{(1)},
b^{(2)},...,b^{(k)}\rangle_{atom}\otimes|n\rangle_{photon}$, then
the mixing angle $\theta$ is adiabatically rotated from $0$ to
$\pi/2$ by keeping the ratio between arbitrary two of the
Rabi-frequencies $\Omega_1$, $\Omega_2$ ... and $\Omega_{k}$ in a
fixed value (i.e. keeping the mixing angles $\phi_j$ constant) and
switching them off adiabatically, we finally obtain the state from
the dark-state of present system:
\begin{eqnarray}\label{eqn:atom-spliter1}
|\Psi(t)\rangle&=&\sum_{n}P_n(\alpha_0)|D_n(\theta=\frac{\pi}{2})\rangle\nonumber\\
&=&\sum_{j}\sum_{l}...\sum_{f}P_j(\alpha_{1})P_l(\alpha_{2})...P_f(\alpha_{k})
|c^{(1)}_j,c^{(2)}_l,...,c^{(k)}_f\rangle\otimes|0\rangle\nonumber\\
&=&|\alpha_1,
\alpha_2,...,\alpha_k\rangle_{coherence}\otimes|0\rangle_{photon},
\end{eqnarray}
where $\alpha_1=\alpha_0\prod_{j=1}^{k-1}\cos\phi_j,
\alpha_l=\alpha_0\sin\phi_{l-1}\prod_{j=l}^{k-1}\cos\phi_j,
(l=2,3,...,k)$. The above expression clearly shows that the
injected quantized field can be stored in the $k$ atomic
ensembles. Particularly, if the strengths of all control fields
keep the same value during the process that they are turned off,
the final atom coherence reads
$\alpha_{1}=\alpha_{2}=...=\alpha_{k}=\alpha_0/\sqrt{k}$, which
means the quantum information of the initial probe light is stored
homogeneously in the $k$ atomic ensembles.

\section{Generation of quantum entanglement}

In above section, we discussed the interesting phenomenon that one
input coherent probe light can convert into many different output
coherent probe lights via the dark-state evolution process. In
this section we shall discuss another novel application to the
generation of entangled states of lights or atomic ensembles with
present DSPs theory in multi-level and multi-ensemble atomic
systems. For this we should use a non-classical input probe light,
for example, a superposition of coherent states
\cite{schrodinger}, a single-photon state, etc.

\subsection{Two-photon entanglement}
The coherent entangled states can be obtained with the quantized
DSPs theory of $four$-level system when the injected quantized
field is in a Sch\"{o}dinger cat state \cite{schrodinger}, e.g.
for the initial total state reads
$|\Psi_0\rangle^{\pm}=\frac{1}{\sqrt{{\cal
N}(\alpha_0)}}|0\rangle\otimes(|\alpha_0\rangle\pm|-\alpha_0\rangle)
\otimes|b\rangle$ where the normalized factor ${\cal
N}_{\pm}(\alpha_0)=2\pm2e^{-2|\alpha_0|^2}$, with the same process
discussed in the section III.A (see eq. (\ref{eqn:split1}), set
$m=4$) we find the injected quantized pulse can evolve into a very
interesting entangled coherent state (ECS) of two output fields
($|\Psi_0\rangle^{\pm}\rightarrow|\Psi_e\rangle^{\pm}$)
\begin{eqnarray}\label{eqn:form}
&\frac{1}{\sqrt{{\cal
N}_{\pm}(\alpha_0)}}|0\rangle\otimes\bigr(|\alpha_0\rangle\pm|-\alpha_0\rangle\bigr)
\otimes|b\rangle=\frac{1}{\sqrt{{\cal
N}_{\pm}(\alpha_0)}}\bigr(|0\rangle\otimes|\alpha_0\rangle\pm|0\rangle\otimes|-\alpha_0\rangle\bigr)
\otimes|b\rangle\longrightarrow\nonumber\\
&\longrightarrow\frac{1}{\sqrt{{\cal
N}_{\pm}(\alpha_0)}}\bigr(\sum_{j}\sum_{k}P_j(\alpha_{e1})P_k(\alpha_{e2})
|b,j,k\rangle\pm\sum_{j}\sum_{k}P_j(-\alpha_{e1})P_k(-\alpha_{e2})
|b,j,k\rangle\bigr).
\end{eqnarray}
The final state in above formula can be rewritten as:
\begin{eqnarray}\label{eqn:entangled1}
|\Psi_e\rangle^{\pm}=\frac{1}{\sqrt{{\cal
N}_{\pm}(\alpha_0)}}\bigr(|\alpha_{e1},\alpha_{e2}
\rangle\pm|-\alpha_{e1},-\alpha_{e2}\rangle\bigr)_{photon}
\otimes|b\rangle.
\end{eqnarray}
If $\phi_e=0$, hence $\alpha_{e1}=\alpha_0$ and $\alpha_{e2}=0$,
and then the evolution of the quantized fields proceed as
$|0\rangle\otimes(|\alpha_0\rangle\pm|-\alpha_0\rangle)/\sqrt{{\cal
N}_{\pm}(\alpha_0)}\rightarrow(|\alpha_0\rangle\pm|-\alpha_0\rangle)\otimes|0\rangle/\sqrt{{\cal
N}_{\pm}(\alpha_0)}$, which means the input Sch\"{o}dinger cat
state is now fully converted into another one with different
vibrational mode. On the other hand, for the general case of
non-zero value of the coherent parameters $\alpha_{e1}$ and
$\alpha_{e2}$, the states of output quantized fields are entangled
coherent states. Since the parameters $\alpha_{ei} (i=1,2)$ is
controllable, the entanglement of the output states
\cite{entanglement} $E^{\pm}(\alpha_{e1}, \alpha_{e2})=-$
tr$(\rho^{\pm}_{\alpha_{e1}}\ln\rho^{\pm}_{\alpha_{e1}})$ with the
reduced density matrix $\rho^{\pm}_{\alpha_{e1}}=$
tr$^{(\alpha_{e2}, atom)}(|\Psi_e\rangle\langle\Psi_e|)^{\pm}$ can
also easily be controlled by the re-applied control fields. In
particular, for the initial state $|\Psi_0\rangle^{-}$, if
$\phi_e=\pi/4$, we have
$\alpha_{e1}=\alpha_{e2}=\alpha_0/\sqrt{2}$ and then obtain the
maximally entangled state(MES):
$|0\rangle\otimes\bigr(|\alpha_0\rangle-|-\alpha_0\rangle\bigr)/\sqrt{{\cal
N}_-(\alpha_0)}\rightarrow\bigr(|\frac{\alpha_0}{\sqrt{2}},\frac{\alpha_0}
{\sqrt{2}}\rangle-|-\frac{\alpha_0}{\sqrt{2}},-\frac{\alpha_0}{\sqrt{2}}\rangle\bigr)/\sqrt{{\cal
N}_-(\alpha_0)}$ which is most useful for quantum information
process. With the definitions
$|+\rangle=\bigr(|\frac{\alpha_0}{\sqrt{2}}\rangle+|-\frac{\alpha_0}{\sqrt{2}}\rangle\bigr)/\sqrt{{\cal
N}_+(\alpha_0/2)}$ and
$|-\rangle=\bigr(|\frac{\alpha_0}{\sqrt{2}}\rangle-|-\frac{\alpha_0}{\sqrt{2}}\rangle\bigr)/\sqrt{{\cal
N}_-(\alpha_0/2)}$, the output state can be rewritten as
$(|+\rangle|-\rangle+|-\rangle|+\rangle)/\sqrt{2}$ which is the
maximum entangled state of output light pulses. Generalization of
these results to multi-mode probe pulses is straightforward. Since
our scheme of generating the entangled coherent states via
quantized DSPs theory is linear and controllable and it only
requires a macroscopic quantum superposition for the initial
state, this scheme deserves study in experiment which has made
much progress recent years \cite{ent}. Remarkably, the latest
works have reported the experimental realization of EIT quantum
memory in three-level system\cite{ent2}. For our scheme, the key
point in experiment is to store the quantum states of one
non-classical probe light in a multi-level system, e.g. a
four-level system, and then use two control fields to convert
quantum states of the initial probe light into an entangled state
of two output pulses. Since the quantum memory for few probe
photons is experimentally realized in three-level system, our
scheme of generating entanglement of photons via multi-level
system may be reached in near future. Also, our scheme is
different from those schemes of generating entangled coherent
states via Kerr effect \cite{entangled} and entanglement swapping
using Bell-state measurement \cite{swap}, which are very important
and have been widely studied.

Consider now a different type of input quantum state corresponding
to a single-photon state, i.e. meanwhile the initial total state
\begin{eqnarray}\label{eqn:initial2}
|\Psi_0\rangle=(|0\rangle\otimes|1\rangle)_{photon}\otimes|b\rangle.
\end{eqnarray}
Similarly, after the light state storage and release process
discussed above, one can easily obtained the final entangled
states of two probe photons:
\begin{eqnarray}\label{eqn:photonentangled3}
\Phi_{photon}=\frac{1}{\sqrt{2}}\bigr(|1\rangle|0\rangle+|0\rangle|1\rangle\bigr)_{photon}.
\end{eqnarray}
Also, if the input quantum state corresponding to a multi-photon
state, we can obtain many other entangled forms of the two output
probe lights.

\subsection{Three-photon entanglement via $five$-level EIT}

Here we consider the similar case that the injected quantized
field is in a Sch\"{o}dinger cat state, e.g., meanwhile from the
eq. (\ref{eqn:split1}) (set $m=5$) the initial total state reads
$|\Psi_0\rangle^{\pm}=\frac{1}{\sqrt{{\cal
N}(\alpha_0)}}|0,0\rangle\otimes(|\alpha_0\rangle\pm|-\alpha_0\rangle)
\otimes|b\rangle$ where the normalized factor ${\cal
N}_{\pm}(\alpha_0)=2\pm2e^{-2|\alpha_0|^2}$, with the similar
process used for two-photon entanglement generation we find the
injected quantized pulse can evolve into the very interesting
entangled coherent states (ECS) of three output fields
($|\Psi_0\rangle^{\pm}\rightarrow|\Psi_e\rangle^{\pm}$)
\begin{eqnarray}\label{eqn:entangled2}
&\frac{1}{\sqrt{{\cal
N}_{\pm}(\alpha_0)}}|0,0\rangle\otimes\bigr(|\alpha_0\rangle\pm|\beta_0\rangle\bigr)
\otimes|b\rangle\rightarrow\nonumber\\
&\longrightarrow\frac{1}{\sqrt{{\cal
N}_{\pm}(\alpha_0)}}\bigr(|\alpha_{e1},\alpha_{e2},\alpha_{e3}
\rangle\pm|\beta_{e1},\beta_{e2},\beta_{e3}\rangle\bigr)_{photon}
\otimes|b\rangle,
\end{eqnarray}
where $\alpha_{e1}=\cos\phi\cos\varphi\alpha_0,
\alpha_{e2}=\sin\phi\cos\varphi\alpha_0,
\alpha_{e3}=\sin\varphi\alpha_0$ and
$\beta_{e1}=\cos\phi\cos\varphi\beta_0,
\beta_{e2}=\sin\phi\cos\varphi\beta_0$ and
$\beta_{e3}=\sin\varphi\beta_0$. If $\phi=\pi/4$ and
$\varphi=\tan^{-1}\frac{\sqrt{2}}{2}$, we get
$\alpha_{ej}=\alpha=\alpha_0/\sqrt{3},
\beta_{ej}=\beta=\beta_0/\sqrt{3} (j=1,2,3)$, and the final state
of the atom coherence: $(|\alpha,\alpha,\alpha
\rangle\pm|\beta,\beta,\beta\rangle)_{photon}/\sqrt{{\cal
N}_{0\pm}}$. With the definitions
$|+\rangle=\bigr(|\frac{\alpha_0}{\sqrt{2}}\rangle+|-\frac{\alpha_0}{\sqrt{2}}\rangle\bigr)/\sqrt{{\cal
N}_+(\alpha_0/\sqrt{3})}$ and
$|-\rangle=\bigr(|\frac{\alpha_0}{\sqrt{2}}\rangle-|-\frac{\alpha_0}{\sqrt{2}}\rangle\bigr)/\sqrt{{\cal
N}_-(\alpha_0/\sqrt{3})}$, the output state can be rewritten as

\begin{eqnarray}\label{eqn:threephoton1}
&\Phi_{photon}(+)=\frac{1}{\sqrt{{\cal
N}_{0+}}}\bigr(|\frac{\alpha_0}{\sqrt{3}},\frac{\alpha_0}{\sqrt{3}},\frac{\alpha_0}{\sqrt{3}}
\rangle+|\frac{\beta_0}{\sqrt{3}},\frac{\beta_0}{\sqrt{3}},\frac{\beta_0}{\sqrt{3}}\rangle\bigr)_{photon}
=h_1|+\rangle|+\rangle|+\rangle+h_2|W_+\rangle
\end{eqnarray}
and
\begin{eqnarray}\label{eqn:threephoton2}
&\Phi_{photon}(-)=\frac{1}{\sqrt{{\cal
N}_{0-}}}\bigr(|\frac{\alpha_0}{\sqrt{3}},\frac{\alpha_0}{\sqrt{3}},\frac{\alpha_0}{\sqrt{3}}
\rangle-|\frac{\beta_0}{\sqrt{3}},\frac{\beta_0}{\sqrt{3}},\frac{\beta_0}{\sqrt{3}}\rangle\bigr)_{photon}=
h'_1|-\rangle|-\rangle|-\rangle+h'_2|W_-\rangle,
\end{eqnarray}
where
$|W_+\rangle=|+\rangle|-\rangle|-\rangle+|-\rangle|+\rangle|-\rangle+|-\rangle|-\rangle|+\rangle$
and
$|W_-\rangle=|-\rangle|+\rangle|+\rangle+|+\rangle|-\rangle|+\rangle+|+\rangle|+\rangle|-\rangle$
are $W$ states \cite{w-state} of three light fields.
$h_1=\sqrt{N_+N^2_-/16N^2_{0+}}$,
$h'_1=\sqrt{N_-N^2_+/16N^2_{0-}}$, $h_2=\sqrt{N^3_+/4N^2_{0+}}$
and $h'_2=\sqrt{N^3_-/16N^2_{0-}}$. The
eqs.(\ref{eqn:threephoton1}) and (\ref{eqn:threephoton2}) indicate
a fascinating phenomenon: The $two$-light state is $still$
entangled after reducing the third one. Similar to the result of
eqs. (\ref{eqn:initial2}) and (\ref{eqn:photonentangled3}), when
the input probe light is in a single-photon state, one can obtain
the maximum entangled states of three-mode photons.

It is noteworthy that the five-qubit code entanglement can be
obtained via a seven-level system that interacts with five probe
and five control fields, which is the shortest code that can be a
error correcting code (ECC) \cite{correct}. Furthermore,
theoretically the entanglement of $m$ light fields can be obtained
using the quantized DSPs theory in multi-level atomic system.

\subsection{Entanglement between two and three atomic ensembles}
Generation of entanglement between atomic ensembles has attracted
much attentions in very recent years \cite{ensemble}. Here we also
can generate entanglement between atomic ensembles by using
multi-atomic-ensemble EIT technique, which is similar to that in
generation of entanglement between coherent lights.

Firstly, one can find that if the injected quantized field is in a
Sch\"{o}dinger cat state \cite{schrodinger}, e.g., for the initial
total state reads \cite{atom}
$|\Psi_0\rangle^{\pm}=\frac{1}{\sqrt{{\cal
N}_{\pm}(\alpha_0)}}\bigr(|\alpha_0\rangle\pm|-\alpha_0\rangle\bigr)_{photon}
\otimes|b^{(1)},b^{(2)}\rangle_{atom}$ where the normalized factor
${\cal N}_{\pm}(\alpha_0)=2\pm2e^{-2|\alpha_0|^2}$, with the
scheme discussed above (see eq. (\ref{eqn:atom-spliter1}), set
$k=2$) we can finally obtain a very interesting entangled atomic
coherence of two atomic ensembles
($|\Psi_0\rangle^{\pm}\rightarrow|\Psi_e\rangle^{\pm}$)
\begin{eqnarray}\label{eqn:entangled3}
&\frac{1}{\sqrt{{\cal
N}_{\pm}(\alpha_0)}}\bigr(|\alpha_0\rangle\pm|-\alpha_0\rangle\bigr)_{photon}
\otimes|b^{(1)},b^{(2)}\rangle_{atom}\rightarrow\nonumber\\
&\longrightarrow\frac{1}{\sqrt{{\cal
N}_{\pm}(\alpha_0)}}|0\rangle_{photon}\otimes\bigr(|\alpha_{1},\alpha_{2}
\rangle\pm|-\alpha_{1},-\alpha_{2}\rangle\bigr)_{coherence}.
\end{eqnarray}

Particularly, for the initial state $|\Psi_0\rangle^{-}$, if
$\phi=\pi/4$, we have $\alpha_{1}=\alpha_{2}=\alpha_0/\sqrt{2}$
and then obtain the maximally entangled state
(MES):$(|+\rangle|-\rangle+|-\rangle|+\rangle)_{coherence}/\sqrt{2}$,
where
$|+\rangle=\bigr(|\frac{\alpha_0}{\sqrt{2}}\rangle+|-\frac{\alpha_0}{\sqrt{2}}\rangle\bigr)_{coherence}/\sqrt{{\cal
N}_+(\alpha_0/2)}$ and
$|-\rangle=\bigr(|\frac{\alpha_0}{\sqrt{2}}\rangle-|-\frac{\alpha_0}{\sqrt{2}}\rangle\bigr)_{coherence}/\sqrt{{\cal
N}_-(\alpha_0/2)}$ are the orthogonal basis.

Secondly, the $three$-atomic-ensemble entanglement can easily
obtained for the case of $m=3$. Considering the Sch\"{o}dinger cat
state of the injected probe field, for example, if
$|\Psi_0\rangle^{\pm}=\frac{1}{\sqrt{{\cal
N}_{0\pm}}}\bigr(|\alpha_0\rangle\pm|\beta_0\rangle\bigr)_{photon}
\otimes|b^{(1)},b^{(2)},b^{(3)}\rangle_{atom}$ with the normalized
factor ${\cal N}_{0\pm}=2\pm2e^{-|\alpha_0-\beta_0|^2/2}$, the
entangled quasi spin-waves between $3$-atomic ensembles can be
obtained by properly steering the external control fields
\begin{eqnarray}\label{eqn:threeatom1}
&\frac{1}{\sqrt{{\cal
N}_{0\pm}}}\bigr(|\alpha_0\rangle\pm|\beta_0\rangle\bigr)_{photon}
\otimes|b^{(1)},b^{(2)},b^{(3)}\rangle_{atom}\rightarrow\nonumber\\
&\longrightarrow\frac{1}{\sqrt{{\cal
N}_{0\pm}}}|0\rangle_{photon}\otimes\bigr(|\alpha_{1},\alpha_{2},\alpha_{3}
\rangle\pm|\beta_{1},\beta_{2},\beta_{3}\rangle\bigr)_{coherence},
\end{eqnarray}
where $\alpha_1=\cos\phi\cos\varphi\alpha_0,
\alpha_2=\sin\phi\cos\varphi\alpha_0,
\alpha_3=\sin\varphi\alpha_0$ and
$\beta_1=\cos\phi\cos\varphi\beta_0,
\beta_2=\sin\phi\cos\varphi\beta_0$ and
$\beta_3=\sin\varphi\beta_0$. Similar to the eqs.
(\ref{eqn:threephoton1}) and (\ref{eqn:threephoton2}), the final
entangled states can then be rewritten as
\begin{eqnarray}\label{eqn:threeatom2}
\Phi_{123}(+)=\frac{1}{\sqrt{{\cal N}_{0+}}}(|\alpha,\alpha,\alpha
\rangle+|\beta,\beta,\beta\rangle)_{coherence}=
h_1|+\rangle|+\rangle|+\rangle+h_2|W_+\rangle
\end{eqnarray}
and
\begin{eqnarray}\label{eqn:threeatom3}
\Phi_{123}(-)=\frac{1}{\sqrt{{\cal N}_{0-}}}(|\alpha,\alpha,\alpha
\rangle-|\beta,\beta,\beta\rangle)_{coherence}=
h'_1|-\rangle|-\rangle|-\rangle+h'_2|W_-\rangle,
\end{eqnarray}
where the coefficients $h_{1,2}$ and $h'_{1,2}$ has the same form
as that in eq. (\ref{eqn:threephoton1}) and
(\ref{eqn:threephoton2}), and $|W_{\pm}\rangle$ are the
corresponding $W$ states. Furthermore, theoretically one can
generate entangled atomic states between multi-atomic ensembles by
extending present results to $m$-atomic-ensemble system.

The above results show many similar features between multi-level
systems and multi-ensemble system. In fact, in present large
number atoms and weak excitation case, the collective atomic
operators satisfy the same commutation relations with the photonic
boson operators ($\hat a_j, \hat a_j^{\dag}$). Therefore, we can
readily conclude a general understanding of the processes that a
quantized probe field can be transferred into many probe ones in
multi-level system and can be transferred into many ensembles of
atomic coherence, say, the process can be generally regarded that
a bosonic field can be transferred into many different bosonic
ones via EIT quantum memory technique. This may be the basis that
we can use multi-level system to generate multi-photon
entanglement and use many-ensemble system to generate
multi-atomic-ensemble entanglement.

Before conclusion, we should emphasize again the adiabatic
condition in the EIT quantum memory process with multi-level and
multi-ensemble atomic systems. As we have known, the condition of
adiabatic evolution is most important for the quantum memory
technique based on the quantized DSPs theory, because the total
system should be confined in dark-state subspace during the
process of quantum memory. It is interesting that the symmetrical
properties of the multi-level system and multi-atomic-ensemble
system are dependent on parameters such as the coupling constant
$g_i$ and Rabi frequency $\Omega_i$ etc. For multi-level system,
the largest zero-degeneracy class besides dark-state subspace will
exist for the case $g_1=g_2=...=g_{m-2}$ \cite{liu}, while for the
multi-ensemble atomic system, it will do when
$\Omega_1=\Omega_2=...=\Omega_k$ \cite{atom}. For example, we can
give a brief discussion on the $five$-level system which has the
largest degeneracy class when the parameters satisfy
$g_1=g_2=g_3=g$. For this we define
\begin{eqnarray}\label{eqn:operator3}
\hat u&=&\cos\phi\hat E_1+\sin\phi\hat E_2 , \ \ \
 \hat v=-\sin\phi\hat E_1+\cos\phi\hat E_2 ; \nonumber\\
\hat s&=&\cos\varphi\hat u+\sin\varphi\hat E_3 , \ \ \
 \hat f=-\sin\varphi\hat u+\cos\varphi\hat E_3; \nonumber\\
 \hat a_{12+}&=&\cos\phi\hat a_1+\sin\phi\hat a_2, \ \ \hat
a_{12-}=-\sin\phi\hat a_1+\cos\phi\hat a_2; \nonumber\\
\hat a_{123+}&=&\cos\varphi\hat a_{12+}+\sin\varphi\hat a_3, \ \
\hat a_{123-}=-\sin\varphi\hat a_{12+}+\cos\varphi\hat
a_3\nonumber
\end{eqnarray}
and the BSPs operator $\hat b=\sin\theta\hat
a_{123+}+\cos\theta\hat C$. Using these definitions one can find
the shift operators as follow
\begin{eqnarray}\label{eqn:operator4}
\hat Q_{\pm}^{\dag}=\cos\phi\hat s^{\dag}\pm\sin\phi \ \hat
b^{\dag}, \ \
 \hat P_{\pm}^{\dag}=\hat v^{\dag}\pm\hat a_{12-}^{\dag}, \ \
\hat O_{\pm}^{\dag}=\hat f^{\dag}\pm\hat a_{123-}^{\dag},
\end{eqnarray}
which satisfy the commutation relations $[\hat H, \hat
Q_{\pm}^{\dag}]=\pm\epsilon_1\hat Q_{\pm}^{\dag}, \ \ [\hat H,
\hat P_{\pm}^{\dag}]=\pm\epsilon_2\hat P_{\pm}^{\dag}, \ \ [\hat
H, \hat O_{\pm}^{\dag}]=\pm\epsilon_3\hat O_{\pm}^{\dag}$, where
$\epsilon_1=\sqrt{g^2N+\Omega_1^2+\Omega_2^2+\Omega_3^2}$ and
$\epsilon_2=\epsilon_3=g\sqrt{N}$. Thanks to these results we
finally obtain the largest degeneracy class of present system:
\begin{eqnarray}\label{eqn:degeneracy3}
|r(i,j;k,l;f,g;n)\rangle=\frac{1}{\sqrt{i!j!k!l!f!g!}}(\hat
Q_+^{\dag})^i(\hat Q_-^{\dag})^j(\hat P_+^{\dag})^k(\hat
P_-^{\dag})^l(\hat O_+^{\dag})^f(\hat O_-^{\dag})^g|D_n\rangle
\end{eqnarray}
with eigenvalue
$E(i,j;k,l;f,g)=(i-j)\epsilon_1+[(k+f)-(l+g)]\epsilon_2$. We
notice that for each given pair of indices $(i,j)$ and
$(k+f,l+g)$, $\{|r(i,j;k,l;f,g;n)\rangle \ |n=0,1,2,\cdots \}$
defines a degenerate set of eigenstates. When $i=j$ and
$k+f=l+g=m$, $E(i,i;k+f=l+g)=0$, and a larger zero-eigenvalue
degeneracy class is given by:
$\{|r(i,i;k,l;m-k,m-l;n)\rangle=|d(i,k,l,m;n) \
|m-k\geq0,m-l\geq0;i,k,l,m,n=0,1,2,\cdots\}$, i.e.
\begin{eqnarray}\label{eqn:degeneracy4}
|d(i,k,l,m;n)\rangle=\frac{1}{i!k!}(\hat Q_+^{\dag}\hat
Q_-^{\dag})^i(\hat P_+^{\dag})^k(\hat P_-^{\dag})^l(\hat
O_+^{\dag})^{m-k}(\hat O_-^{\dag})^{m-l}|D_n\rangle \ \
(i,k,n=0,1,2,\cdots),
\end{eqnarray}
which is constructed by acting ($\hat Q_+^{\dag}\hat Q_-^{\dag}$)
$i$ times, $\hat P_+^{\dag}$ $k$ times, $\hat P_-^{\dag}$ $l$
times $\hat O_+^{\dag}$ $m-k$ times and $\hat O_-^{\dag}$ $m-l$
times on $|D_n\rangle$. Only when $i=k=l=m=0$, the larger
degeneracy
class reduces to the special dark-state subset $%
\{|D_n\rangle \ |$$ n=0,1,2,\cdots \}$ of present $five$-level
atomic system. However, following the method developed in Refs.
\cite{8,liu,atom} it is straightforward to confirm that any
transition from dark states to other zero-eigenvalue subspace is
also forbidden and therefore the robustness of present general EIT
quantum memory technique is still perfect, even in the large
zero-degeneracy case.

\section{Conclusions and further discussions}

To sum up, the single-ensemble composed of multi-level atoms and
multi-ensemble composed of $three$-level atoms with EIT are
studied in detail in this paper, focused on the interesting
dynamical symmetry and its applications to quantum information
processing. The general definition of dark-state polaritons
(DSPs), and then the dark-states of these different systems are
obtained by discovering the symmetrical Lie group of various
atomic systems, such as single-atomic-ensemble composed of complex
$m$-level $(m>3, multi-level)$ atoms, and multi-atomic-ensemble
system composed of of $three$-level atoms. It is interesting that
the symmetrical properties of the multi-level system and
multi-atomic-ensemble system are dependent on some characteristic
parameters of the EIT system. Furthermore, a controllable scheme
to generate quantum entanglement between light fields or different
atomic ensembles via quantized DSPs theory is discussed, which
might be experimentally implemented in the near future..

It is noteworthy that there are many counterparts between the
multi-level (single-ensemble) and multi-ensemble atomic systems.
For example, the entanglement between two light fields (or among
three light fields) can be generated using $four$- (or $five$-)
level system, and the entanglement between two (or among three)
ensembles of atoms can be generated via $two$- (or $three$-)
atomic-ensemble system; The dynamical symmetry of $four$-level
system is governed by the Lie group
$SU(3)\overline{\otimes}H_{3}$, while that of
$two$-atomic-ensemble system is partly governed by
$SO(4)\overline{\otimes}H_{4}$; The symmetrical properties of
multi-level system is dependent on the parameter of coupling
constant $g_i$ of probe fields while that of multi-atomic-ensemble
system is dependent on the Rabi-frequency of control fields, and
the larger degeneracy class of multi-level system is just similar
to that of the corresponding multi-atomic-ensemble system, etc..
All these interesting aspects may deserve further study in next
work.

\bigskip

\noindent

We thank professors Yong-Shi Wu and J. L. Birman for valuable
discussions. This work is supported by NSF of China under grants
No.10275036 and No. 10304020, and by Wuhan open fund of state key
laboratory of magnetic resonance and atomic and molecular physics,
No. T152505.




\bigskip

\noindent


\begin{thebibliography}{99}
\bibitem{2} S. E. Harris, J. E. Field and A. Kasapi, Phys. Rev. A 46, R29 (1992);
            M. O. Scully and M. S. Zubairy, Quantum Optics (Cambridge University Press, Cambridge 1999).
\bibitem{3} L. V. Hau et al., Nature (London) 397, 594 (1999); M. M. Kash et al., Phys.Rev.Lett.82, 529(1999);
            C. Liu, Z. Dutton, C. H. Behroozi and L. V. Hau, Nature (London) 409, 490 (2001);
            D. F. Phillips, A. Fleischhauer, A. Mair, R. L. Walsworth and M. D. Lukin, Phys. Rev. Lett. 86,783(2001);
\bibitem{4} M. D. Lukin and A. Imamo\v{g}lu, Phys. Rev. Lett. 84, 1419 (2000);
            M. D. Lukin, S. F. Yelin and M. Fleischhauer, Phys. Rev. Lett. 84, 4232 (2000);
            M. Fleischhauer and S. Q. Gong, Phys. Rev. Lett. 88, 070404 (2002);
            C. Mewes and M. Fleischhauer, Phys. Rev. A 66, 033820 (2002).
\bibitem{5} Y. Wu, J. Saldana and Y. Zhu, Phys. Rev. A 67, 013811 (2003);
            Y. Li, P. Zhang, P. Zanardi and C. P. Sun, quant-ph/0402177 (2004);
            G.Juzeli\={u}nas and P.\"{O}hberg, Phys.Rev.Lett. 93, 033602(2004);
            L. M. Kuang and L. Zhou, Phys. Rev. A 68, 043606 (2003);
            X. J. Liu, H. Jing and M. L. Ge, Phys. Rev. A 70, 055802 (2004).
\bibitem{6} M. Fleischhauer and M. D. Lukin, Phys. Rev. Lett. 84, 5094 (2000);
            M. Fleischhauer and M. D. Lukin, Phys. Rev. A 65, 022314 (2002).
\bibitem{7} M. D. Lukin, Rev. Mod. Phys. 75, 457 (2003).
\bibitem{Lidar} Irreversible Quantum Dynamics, Edited by F. Benatti and R. Floreanini, "Decoherece-Free subspace
                and Subsystem" by D. A. Lidar and B. Whaley, pp. 83-120, Springer Lecture Notes in Physics vol.622, Berlin,
                (2003); e-print:quant-ph/0301032(2003).
\bibitem{8} C. P. Sun, Y. Li and X. F. Liu, Phys. Rev. Lett. 91, 147903 (2003).
\bibitem{liu}  X. J. Liu, H. Jing, X. T. Zhou and M. L. Ge, Phys. Rev. A, 70,015603(2004);
               X. J. Liu, H. Jing and M. L. Ge, quant-ph/0403171.
\bibitem{atom} H. Jing, X. J. Liu, M. L. Ge and M. S. Zhan, Phys. Rev. A 71, 062336 (2005).
\bibitem{dick} R. H. Dick, Phys. Rev. 93, 99 (1954).
\bibitem{group} B. G. Wybourne, {\it Classical Groups for Physicists} (John Wiley, NY, 1974); M. A. Shifman,
             {\it Particle Physics and Field Theory}, p775 (World Scientific, Singapore,
             1999).
\bibitem{gell} F. T. Hioe, Phys. Rev. A, 32, 2824 (1985); Phys. Rev. A 28, 879 (1983).
\bibitem{four1} A. B. Matsko, et al., At. Mol. Opt. Phys. 46, 191 (2001);
                A. S. Zibrov et al., Phys. Rev. Lett. 88, 103601 (2002).
\bibitem{four2} A. Andr\'{e} and M. D. Lukin, Phys. Rev. Lett. 89, 143602 (2002));
                A. Raczy\'{n}ski and J. Zaremba, Opt. Commun. 209, 149 (2002); quant-ph/0307223 (2003).
\bibitem{ground} M. D. Lukin, S. F. Yelin and M. Fleischhauer, Phys. Rev. Lett., 84, 4232 (2000);
                 E. Arimondo, Progr. In Optics 35, 259 (1996).
\bibitem{schrodinger} J. J. Slosser and G. J. Milburn, Phys. Rev. Lett. 75, 418(1995).
\bibitem{entanglement} O. Hirota, quant-ph/0101096(2001).
\bibitem{ent} D. N. Matsukevich and A. Kuzmich, Science, 306, 663 (2004);
              C. H. van der Wal et al., Science, 301, 196 (2003); O. Mandel et al., Science, 425, 937 (2003);
              K. Hammerer et al., arXiv: quant-ph/0312156 (2003).
\bibitem{ent2} T. Chaneli\`{e}re et al., Nature 438, 833 (2005); M. D. Eisaman et al., Nature 438, 837 (2005).
\bibitem{entangled} M. Paternostro, M. S. Kim, and B. S. Ham, Phys. Rev. A 67, 023811 (2003);
\bibitem{swap} Xiaoguang Wang and Barry C. Sanders, Phys. Rev. A 65, 012303 (2003);
               F. L. Kien et al., Phys. Rev. A 68, 063803 (2003);
               N. A. Ansari et al., Phys. Rev. A 50, 1492 (1994).
\bibitem{w-state} V. Coffman, J. Kundu and W. K. Wootters, Phys. Rev. A 61, 052306 (2000);
                  M. A. Nielsen and I. Chuang, Quantum Computation and Quantum Information
                  (Cambridge University Press, Cambridge, 2000).
\bibitem{correct} C. H. Bennett, D. P. DiVincenzo, J. A. Smolin, and W. K. Wootters, Phys. Rev. A 54, 3824 (1996);
                  R. Laflamme, C. Miquel, J.-P. Paz, and W. H. Zurek, Phys. Rev. Lett. 77, 198 (1996).
\bibitem{ensemble} A. Dantan, A. Bramati and M. Pinard, Europhys. Lett. 67, 881 (2004);
                  V. Josse, A. Dantan, A. Bramati, M. Pinard and E. Giacobino, Phys. Rev. Lett. 92, 123601 (2004).

\end{thebibliography}
\end{document}